\documentclass[final,3p,times]{elsarticle}
\usepackage{graphicx}
\usepackage{amssymb}
\biboptions{comma,sort&compress}

\journal{Physica A}
\begin{document}
\begin{frontmatter}


\title{Stochastic resonance in an RF SQUID with shunted ScS junction}
\author[ilt]{O.G. Turutanov\corref{cor1}}
\ead{turutanov@ilt.kharkov.ua} \cortext[cor1]{Corresponding author.}
\author[pert]{V.A. Golovanevskiy}
\author[ilt]{V.Yu. Lyakhno}
\author[ilt]{V.I. Shnyrkov}
\address[ilt]{B.Verkin Institute for Low Temperature Physics and Engineering, NAS Ukraine, 47 Lenin Ave., Kharkov 61103, Ukraine}
\address[pert]{Western Australian School of Mines, Curtin University, Kent St., Bentley, Perth Western Australia 6845}

\begin{abstract}
     Using a point (superconductor-constriction-superconductor, ScS) contact in a
single-Josephson-junction superconducting quantum interference
device (RF SQUID) provides stochastic resonance conditions at any
arbitrary small value of loop inductance and contact critical
current, unlike SQUIDs with more traditional tunnel
(superconductor-insulator-superconductor, SIS) junctions. This is
due to the unusual potential energy of the ScS RF SQUID which always
has a barrier between two wells thus making the device bistable.
This paper presents the results of a numerical simulation of the
stochastic dynamics of the magnetic flux in an ScS RF SQUID loop
affected by band-limited white Gaussian noise and low-frequency sine
signals of small and moderate amplitudes. The difference in stochastic
amplification of RF SQUID loops incorporating ScS and SIS
junctions is discussed.
\end{abstract}

\end{frontmatter}

\section{Introduction}
\label{intro}

The sensitivity of superconducting quantum interference devices
(SQUIDs) and their quantum analogues, SQUBIDs, has practically reached
the quantum limitation \cite{Ketch82,Soroka12,Korolev}. However,
with increase of the quantizing loop inductance up to  $L\sim
10^{-9} -10^{-10} $  H, thermodynamic fluctuations lead to quick
deterioration of the energy resolution. As shown earlier
\cite{Rouse95,Hibbs95,Hibbs98,Turutan02,Gluhov06}, the sensitivity
of magnetometers can be enhanced in this case by using stochastic
resonance (SR). The SR phenomenon whose concept was introduced in the early 1980s
\cite{Benzi81,Eckmann,Nicolis} manifests itself in
non-monotonic rise of a system response to a weak periodic signal
when noise of a certain intensity is added to the system. Owing
to extensive studies during the last two decades, the stochastic resonance
effect has been revealed in a variety of natural and artificial
systems, both classical and quantum. Analytical approaches and
quantifying criteria for estimation of the ordering due to the
noise impact were determined and described in the reviews
\cite{Gamma98,Anish99,Wel04}. In particular, the sensitivity of a
bistable stochastic system fed with a weak periodic signal can be
significantly improved in the presence of thermodynamic or external
noise that provides switching between the metastable states of the system.
For example, it was experimentally proved \cite{Rouse95} that the
gain of a harmonic informational signal can reach 40 dB at a certain
optimal noise intensity in a SQUID with an SIS
(superconductor-insulator-superconductor) Josephson junction.
Moreover, the stochastic amplification in SIS-based SQUIDs can be
maximized at a noise level insufficient to enter the SR mode by
means of the stochastic-parametric resonance (SPR) effect
\cite{Turutan08} emerging in the system due to the combined action of
the noise, a high-frequency electromagnetic field and the weak
informational signal. An alternative way of enhancing the RF SQUID sensitivity is to suppress the noise with  strong (suprathreshold) periodic RF pumping of properly chosen frequency which results in a better signal-to-noise ratio in the output signal \cite{Pankratov1}. In the latter case the switching between metastable states is mainly due to strong regular RF pumping \cite{Pankratov2} unlike SR where the dominating switching mechanism is the joint effect of noise and weak periodic signal \cite{Gamma98,Anish99,Wel04}.

In recent years quantum point contacts (QPCs) with direct conductance
have attracted strong interest from the point of view of both quantum
channel conductance studies and building qubits with high energy
level splitting. Currently, two types of point contacts are
distinguished, depending on the ratio between the contact dimension
$d$  and the electron wave length  $\lambda _{F}=h/p_{F}$ :
$d>>\lambda _{F}$  for a classical point contact \cite{Kulik78} and
$d\sim \lambda _{F}$ for a quantum point contact
\cite{Agrait,Beenakker,Beenakker_arxiv}. Practically,
superconducting QPCs are superconductor-constriction-superconductor
(ScS) contacts of atomic-size (ASCs). The critical currents of such
contacts can take discrete values. The relation $I_{s}^{ScS}
(\varphi )$ between the supercurrent $I_{s}^{ScS} $  and the order
parameter phase $\varphi$ in both classical and quantum cases at
lowest temperatures ($T\to 0$) essentially differs
\cite{Kulik78,Beenakker,Beenakker_arxiv} from the current-phase
relation for an SIS junction described by the well-known Josephson
formula $I_{s}^{SIS} =I_{c} \sin \varphi$. The corresponding
potential energies in the motion equations are therefore different as
well.

When an SIS junction is incorporated into a superconducting loop with
external magnetic flux $\Phi_{e}=\Phi _{0} /2$  (where $\Phi
_{0}=h/2e\approx 2.07\cdot 10^{-15} $  Wb is the magnetic flux quantum)
piercing the loop, its current-phase relation $I_{s}^{SIS} (\varphi
)$ leads to the formation of a symmetric two-well potential energy
$U^{SIS} (\Phi )$ of the whole loop that principally enables the SR
dynamics only for $\beta _{L} =2\pi LI_{c} /\Phi _{0} >1$.
$\beta_{L}$ is a dimensionless non-linearity parameter sometimes
called the main SQUID parameter. In contrast, the potential energy
$U^{ScS} (\Phi )$ of a superconducting loop with a QPC always has a
barrier with a singularity at its top, and two metastable current
states of the loop differing by internal magnetic fluxes  $\Phi $ can
be formally achieved at any vanishingly low $\beta _{L}<<1$. In the
quantum case, the most important consequences of the "singular"
barrier shape are the essential rise of macroscopic quantum
tunneling rate and the increased energy level splitting in flux
qubits \cite{Soroka12,Korolev}.

     In the classical limit, the SR dynamics of a superconducting loop
with ScS Josephson contact and non-trivial potential  $U^{ScS} (\Phi
)$  would differ substantially from the previously explored
\cite{Rouse95,Hibbs95,Hibbs98,Gluhov06} case of the SIS junction and
would be much like the 4-terminal SQUID dynamics \cite{Turutan02}.
In the present work a numerical analysis is given of stochastic
amplification of weak low-frequency harmonic signals in a
superconducting loop broken by an ScS Josephson junction at low
temperatures  $T<<T_{c} $ . Specific focus is given to low critical
currents, i.e. rather high-impedance contacts (ASCs) when  $\beta
_{L} =2\pi LI_{c} /\Phi _{0} <1$.

\section{ScS junction loop model and numerical computation technique}
\label{model}

The stochastic dynamics of the magnetic flux in an RF SQUID loop (inset
in Fig.\,\ref{fig1}a) was studied by numerical solution of the motion
equation (Langevin equation) in the resistively shunted junction
(RSJ) model \cite{Barone}:

\begin{figure}[h]
  \centering
  \includegraphics[width = 0.7\columnwidth]{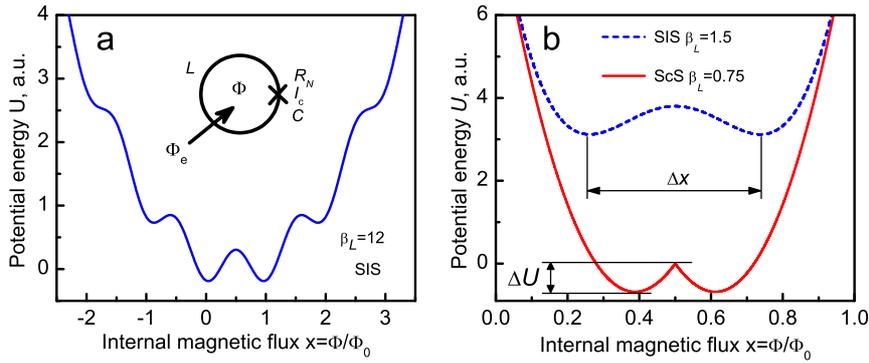}\\
  \caption{(Color online) (a) Potential energy $U^{SIS}$ of an
  SIS-junction-based RF SQUID loop with large non-linearity
  parameter $\beta_{L}=12$ versus the normalized internal magnetic flux
  $x$. The inset is the RF SQUID loop schematic. (b) Potential energies
  of RF SQUIDs with an SIS junction ($\beta_{L}=1.5$) and an ScS junction
  ($\beta_{L}=0.75$) vs. normalized internal magnetic flux $x$. The
  barrier heights $\Delta U$ in both SQUIDs are approximately equal at chosen
  values of $\beta_{L}$. A fixed magnetic flux $\Phi_{e}=\Phi_{0}/2$
  $(x_{e}=1/2)$ is applied to symmetrize the potential.}
  \label{fig1}
\end{figure}

\begin{equation}
LC\frac{d^{2} \Phi (t)}{dt^{2} } +\frac{L}{R} \frac{d\Phi (t)}{dt}
+L\frac{\partial U(\Phi,\Phi_{e} )}{\partial \Phi } =\Phi _{e} (t),
\label{1}
\end{equation}

\noindent
where $C$ is the capacitance; $R$ is the normal shunt resistance of the Josephson junction; $L$ is the loop inductance; $\Phi (t)$  is the internal magnetic flux
in the loop;  $U(\Phi ,\Phi _{e} )$  is the loop potential energy,
which is the sum $U(\Phi ,\Phi _{e} )=U_{M} +U_{J} $ of magnetic
energy of the loop and the coupling energy of the Josephson
junction. The time-dependent external magnetic flux  $\Phi _{e} (t)$
piercing the loop contains a constant and a variable, including noise,
component. This equation is analogous to the motion equation for a
particle of mass $C$ moving in potential $U$  with friction
coefficient $\gamma =1/R$. The junction coupling energy  $U_{J} $ is
specific to its nature; we will consider the case of clean ScS
contacts in the ballistic mode of the electron fly-through
\cite{Kulik78}.

For both classical \cite{Kulik78} and quantum
\cite{Agrait,Beenakker,Beenakker_arxiv} ScS point contacts with the
critical current $I_{c} $, at arbitrary temperature  $T$  the
current-phase relation reads

\begin{equation}
I_{s}^{ScS} (\varphi )=I_{c} \sin \frac{\varphi }{2}\, \tanh
\frac{\Delta (T)\cos \frac{\varphi }{2}}{2k_{B} T} ,\; \; I_{c}
(T)=\frac{\pi \Delta (T)}{eR}, \label{2}
\end{equation}

\noindent
where  $I_{s}^{ScS} (\varphi )$  is the supercurrent
through the contact,  $\Delta (T)$  is the superconducting energy
gap (order parameter),  $\varphi $  is the difference between the
order parameter phases at the contact "banks",  $k_{B} $  is the
Boltzmann constant, $e$  is the electron charge,  and $R$  is the normal
contact resistance. In the limit  $T=0$ the expression (\ref{2})
transforms into

\begin{equation}
I_{s}^{ScS} (\varphi )=I_{c} \sin \frac{\varphi }{2} \;
\textrm{sgn}\,(\cos \frac{\varphi }{2} ) \label{3}
\end{equation}

     The potential energy of a superconducting loop broken by an
ScS contact,  $U^{ScS} (\Phi ,\Phi _{e} )$, reads as

\begin{equation}
U^{ScS} (\Phi ,\Phi _{e} )=\frac{(\Phi -\Phi _{e} )^{2} }{2L}
-E_{J}^{ScS} \left| \cos\frac{\pi \Phi }{\Phi _{0} } \right| ,
\label{4}
\end{equation}

\noindent
where  $E_{J}^{ScS} =I_{c} \Phi _{0} /\pi $  is the maximum
coupling energy of the ScS Josephson contact.

     To compare, the potential energy of a loop with a tunnel
junction is \cite{Barone}

\begin{equation}
U^{SIS} (\Phi ,\Phi _{e} )=\frac{(\Phi -\Phi _{e} )^{2} }{2L}
-E_{J}^{SIS} \cos \frac{2\pi \Phi }{\Phi _{0} }, \label{5}
\end{equation}

\noindent
where  $E_{J}^{SIS} =I_{c} \Phi _{0} /2\pi $  is the maximum
coupling energy of the tunnel Josephson junction.

Reducing the fluxes by the flux quantum  $\Phi _{0} $ :  $x=\Phi /\Phi
_{0} $, $x_{e} =\Phi _{e} /\Phi _{0} $ and the potential energy by
$\Phi _{0}^{2} /2L$, and using the parameter $\beta_{L}$, Eqs. (\ref{4}) and (\ref{5}), can correspondingly be rewritten as

\begin{equation}
\label{6}
u^{ScS} (x,x_{e} )=\frac{(x-x_{e} )^{2} }{2} -\frac{\beta
_{L} }{2\pi ^{2} } \left| \cos \pi x \right|
\end{equation}

\noindent
and

\begin{equation}
\label{7}
u^{SIS} (x,x_{e} )=\frac{(x-x_{e} )^{2} }{2} -\frac{\beta
_{L} }{4\pi ^{2} } \cos (2\pi x)
\end{equation}

     The reduced potential energy  $u^{SIS} (x,\; x_{e} )$  of the loop
with a tunnel junction has two or more local minima at  $\beta _{L}
>1$  only. When the loop is biased by a fixed magnetic flux   $\Phi
_{e} =\Phi _{0} /2$  ($x_{e} =1/2$), the two lowest minima become
symmetric. This case is illustrated in Fig.\,\ref{fig1}a for a large
value  $\beta _{L} =12$, for better illustration.

\begin{figure}[!ht]
  \centering
  \includegraphics[width = 0.7\columnwidth]{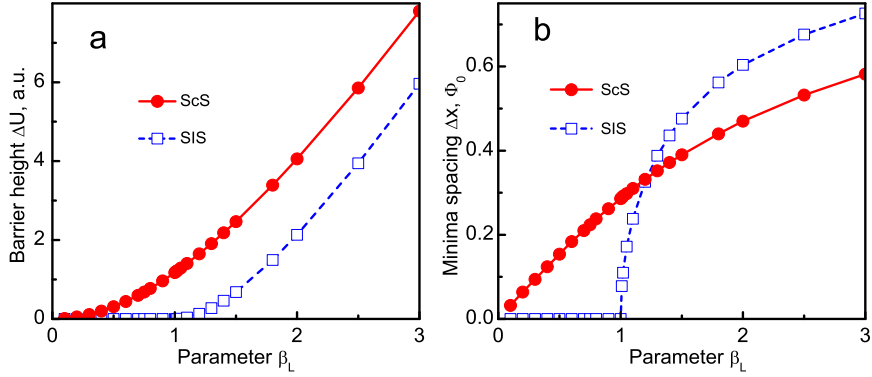}\\
  \caption{(Color online) (a) Energy barrier height $\Delta U$ and
  (b) spacing $\Delta x$ between the potential energy minima versus the
  parameter $\beta_{L}$ for RF SQUIDs with ScS and SIS Josephson
  junctions.} \label{fig2}
\end{figure}

     The essential feature attributed to the potential energy
$u^{ScS} (x,x_{e} )$  of the RF SQUID with ScS contact is that the
inter-well barrier with the singularity at its top keeps its finite
height down to vanishingly small  $\beta _{L} $ and therefore small
$L$ and $I_{c} $. Fig.\,\ref{fig1}b shows the two-well potential of
an RF SQUID with an ScS contact at  $\beta _{L} =0.75<1$  (solid line)
and, for comparison, the potential of the loop with the SIS junction
(dashed line) with the same energy barrier height $\Delta U$ (see
also Fig.\,\ref{fig2}a). Noise of thermal or any other origin
causes switching between the metastable states corresponding to the
minima of  $U(\Phi )$. The average switching rate $r_{sw}$ (of a transition from a metastable state to another one) for white Gaussian noise with intensity $D$  and high barriers ($\Delta U/D \gg 1$) is estimated by the well-known Kramers rate $r_K$ \cite{Kramers}

\begin{equation}
r_{sw}^{smooth} =r_{K} = \frac {\omega_0 \, \omega_b} {2\pi \gamma} \, \exp (-\Delta U/D) \label{8}
\end{equation}

\noindent
for parabola wells and smooth parabola barrier, which is almost the case for the SIS-SQUID potential. Here
$\omega_0 =[U''_{\Phi} (x_{bottom}) / C]^{1/2}$  and $\omega_b =[U''_{\Phi} (x_{top})/C]^{1/2}$  are the angular frequencies of small-amplitude oscillations near the bottom of the well and the top of the barrier, correspondingly, defined by the potential curvature in these points; $\gamma$  is the damping constant.

Meanwhile, for parabola wells and a sharp barrier that is close to the ScS SQUID potential shape, especially at low non-linearity parameter $\beta_L$, the switching time is given by formula (5.4) in \cite{Malakhov}, which in our terms will read as

\begin{equation}
r_{sw}^{sharp} =\sqrt {\pi} \, \frac {\gamma (\Delta x)^2}{8 \Delta U} \sqrt { \frac {D} {\Delta U}} \, \exp (- \Delta U / D) \label{8b}
\end{equation}

     For the thermal noise,  $D=k_{B} T$. In this work we do not
presume any specific nature of the noise, however, considering it
white Gaussian. The sole limitation we impose is setting an upper
cut-off frequency $f_{c} $  for the noise band which does not exceed
the reversal time of the flux relaxation in the loop $1/\tau _{L}
=R/L$ to provide the adiabatic mode for the SQUID operation.
Previous estimations \cite{Gluhov06,Turutan08} following from the
numerical simulation show that a "reasonable" value for $f_{c} $ can
be chosen so that its further increase does not practically affect
the results of the calculations. Usually  $f_{c} \sim (10^{3}
-10^{4} )f_{s} $ is high enough where $f_{s} $  is the signal
frequency. Adding small periodic signal with frequency $f_{s} $  to the
external flux  $\Phi _{e} $ on the noise background enables
stochastic resonance dynamics of a particle in the bistable
potential when the SR condition fulfils

\begin{equation}
r_{sw} \approx 2f_{s} \label{9}
\end{equation}

     For typical experimental parameters, $L\approx 3\cdot 10^{-10}
$  H,  $C\approx 3\cdot 10^{-15} $  F,  $R\approx 1-10^{2} $  Ohm,
$I_{c} \approx 10^{-5} -10^{-6} $ A and  $\beta _{L} =0.1-3$, we
estimate the McCumber parameter accounting for the capacitance to be low
enough: $\beta _{C} =2\pi R^{2} I_{c} C/\Phi _{0} <1$. In this case
(aperiodic, or overdamped, oscillator) the motion is
non-oscillatory, and therefore the first term with second derivative
in Equation \ref{1} can be neglected. Note that a contact with
resistance  $R\sim 100$  Ohm is close to ASC since the number of
conducting channels (atomic chains) is small but the considered
situation, even at low temperatures, remains a classical one because
of strong dissipation. The low signal frequency $f_{s}\sim
1-10\;{\rm Hz}<<1/\tau _{L} $  and the upper-limited noise frequency
band (quasi-white noise) with cut-off frequency $f_{c} \sim 10^{4}
\; {\rm Hz}<<1/\tau _{L} $ make the problem adiabatic, as noted above, and allow one to attribute all the time dependence to the potential energy in Equation (\ref{1}):

\begin{equation}
\tau _{L} \frac{dx}{dt} +\frac{\partial U(x,t)}{\partial x} =0
\label{10}
\end{equation}

     For the case of an ScS contact, by substituting Equation
(\ref{6}) in Equation (\ref{10}), we get

\begin{equation}
\frac{dx}{dt} =\frac{1}{\tau _{L} } \{x_{e} (t)-x+\frac{\beta _{L}
}{2\pi } \sin (\pi x)\cdot sgn [\cos(\pi x)]\}, \label{11}
\end{equation}

\noindent and for an SIS junction, taking into account Equation (\ref{7}),
Equation (\ref{10}) reads as

\begin{equation}
\frac{dx}{dt} =\frac{1}{\tau _{L} } [x_{e} (t)-x+\frac{\beta _{L}
}{2\pi } \sin (2\pi x)] \label{12}
\end{equation}

     The external magnetic flux  $x_{e} (t)$  is the sum of the
fixed bias flux  $x_{dc} =0.5$, the useful signal  $x_{ac} =a\sin
2\pi f_{s} t$  and the noise flux  $x_{N} $. Theoretically, the
noise is assumed to be  $\delta $ -correlated,
Gaussian-distributed, white noise:  $x_{N} =\xi (t)$ , $\left\langle
\xi (t)\xi (t-t')\right\rangle =2D\delta (t-t')$. During numerical
simulation it is emulated by a random-number generator with Gaussian
distribution, variance  $D=\sigma ^{2} $ and repetition period of
about  $2^{90} $. When solving the equation in a finite-difference
approximation, the sampling frequency is  $2^{16} $  which is
equivalent to a noise frequency band of $\sim$ 32 kHz. This allows us to
consider the noise to be quasi-white for stochastic amplification of
the signals with frequency  $f_{s} =1-10$  Hz.

Equations (\ref{11}) and (\ref{12}) were solved
by the Heun algorithm modified for stochastic equations
\cite{Garcia,Kloeden}. 10 to 50 runs were made to obtain 16-second
time series with different noise realizations. They then underwent
fast Fourier transform (FFT), and the resulting spectral densities
$S_{\Phi } (\omega )$  of the output signal (internal flux in the loop)
were averaged. In this work we use the spectral amplitude gain of
the weak periodic signal as the SR quantifier defined as the
ratio of spectral densities of the output and input magnetic fluxes:

\begin{equation}
k(\omega )=S_{\Phi out}^{1/2} (\omega )/S_{\Phi in}^{1/2} (\omega )
\label{13}
\end{equation}

\section{Numerical simulation results and discussion}
\label{result}

The energy barrier height  $\Delta U$, as follows from Equations
(\ref{6}) and (\ref{7}), is determined by  $\beta _{L} $ and is
different for the cases of ScS and SIS junctions
(Fig.\,\ref{fig2}a). As can be seen, in the loop with SIS junction
(referred to as SIS SQUID) the two-well potential with two
metastable states needed to prepare conditions for stochastic
amplification of a weak information signal exists only at  $\beta
_{L} >1$ while it is finite for any $\beta _{L} $  in the ScS SQUID.
Both  $\Delta U$ and $D$ , being in exponent, are the core
parameters to define the switching rate $r_{sw} $ (\ref{8}), (\ref{8b}). For a
specified frequency of a weak harmonic signal, the SR condition (\ref{9}) requirement can be met by increasing the noise power. Meanwhile, the amplitude gain $k(\omega )$ of the small
signal, according to the two-state theory \cite{MCNamara}, should depend
on the spacing $\Delta x$  between the local minima of the potential energy $U(x)$.

\begin{equation}
k(\omega)=\frac{r_{sw} \, (\Delta x)^2}{2 D \, (4 r_{sw}^2+\omega^2)^{1/2}}
\label{14}
\end{equation}

     Fig.\,\ref{fig2}b shows  $\Delta x$  as a function of  $\beta
_{L} $  for the ScS and SIS SQUIDs. It is obvious from
Fig.\,\ref{fig2}b that both the spacing  $\Delta x^{ScS} $  between
the potential energy minima and the barrier height  $\Delta U^{ScS}
$  tend to zero remaining finite when $\beta _{L} \to 0$. In
contrast, for SIS SQUIDs $\Delta U^{SIS}$ and $\Delta x^{SIS}$
vanish at $\beta _{L} =1$.

\begin{figure}[!ht]
  \centering
  \includegraphics[width = 0.7\columnwidth]{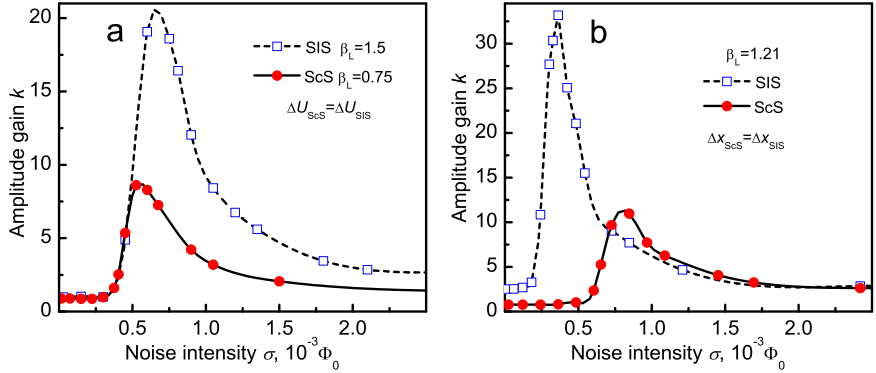}\\
  \caption{(Color online) The amplitude gain $k$ of the sine signal in RF SQUIDs with
  ScS and SIS junctions versus the noise amplitude $\sigma =D^{1/2}$. The $\beta_{L}$ parameters are chosen so that (a) the potential barriers $\Delta U$ in both SQUIDs are equal; (b) the minima spacings $\Delta x$ in both SQUIDs are equal. The signal amplitude $a=0.001$ and the frequency $f_{s}=10$ Hz}.
  \label{fig3}
\end{figure}

     Calculation of the small-signal gain with the same barrier
height for both potentials,  $\Delta U^{ScS} =\Delta U^{SIS} $ ,
shows that maximal gain for an SIS SQUID is roughly two times higher than that
of an SR amplifier based on an ScS SQUID (Fig.\,\ref{fig3}a).

Maximal gain is obtained when the SR condition (\ref{9}) is met. After
substituting (\ref{9}) in (\ref{14}) the gain becomes a function of
only  $\Delta x$ and  $D$. However, the obtained difference in the
gain is less than could be derived from only the ratio of $\Delta
x^{ScS} $  to $\Delta x^{SIS} $ because the gain maxima correspond
to different optimal noise intensities  $D_{m} =\sigma _{m}^{2} $
which depend on the potential shapes modifying the switching rate
$r_{sw}$.
Using $\sigma_m$ from Fig.\ref{fig3}a to calculate the gain ratio by the formula (\ref{14}), we get $k^{SIS}/ \, k^{ScS}=2.15$ versus the experimental value of 2.37. This is good enough taking into account the simplicity of the two-state model. Fig.\ref{fig3}b illustrates the alternative case when the minima spacings for both potentials are equal while the barriers are different. Unexpectedly, there is no agreement here between the simulated and calculated gain ratios.
Nevertheless, it should be stressed that despite the lower
gain in the ScS SQUID, SR amplification in it is possible at very
small critical currents (typical for ASCs) and low noise level (which
may correspond to thermodynamic fluctuations at ultralow
temperatures). Meanwhile, there is no amplification of weak
informational signals in SIS SQUIDs for all $\beta _{L} <1$.

\begin{figure}[h]
  \centering
  \includegraphics[width = 0.7\columnwidth]{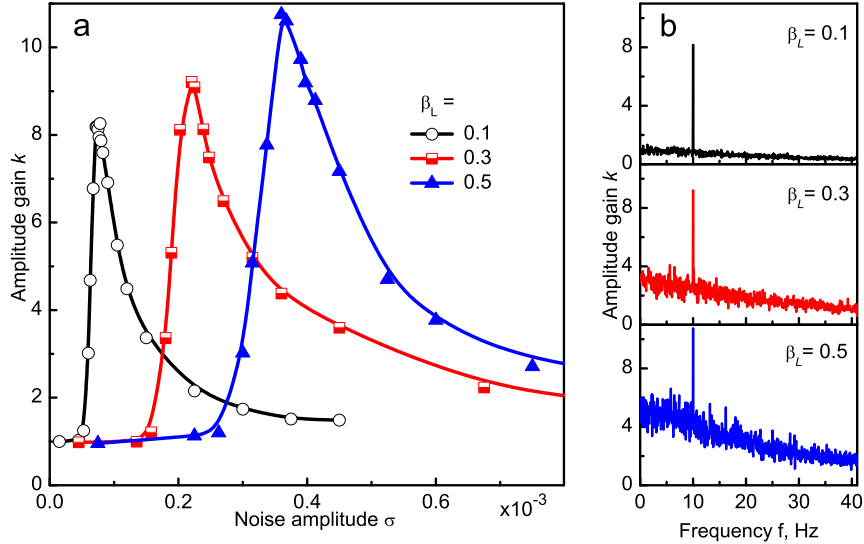}
  \caption{(Color online) (a) The amplitude gain $k$ of the sine signal in an RF SQUID with ScS junction for various
  $\beta_{L}=0.1,\, 0.3,\, 0.5$ versus the noise amplitude
  $\sigma=D^{1/2}$. (b) Spectral gains $k(f)$ for the same values of
  $\beta_{L}$ as in panel (a) and noise levels corresponding to the
  peak of each curve in panel (a).  The signal amplitude $a=0.001$ and the
  frequency $f_{s}=10$ Hz.} \label{fig4}
\end{figure}

     Fig.\,\ref{fig4} displays a set of SR gain in ScS SQUID versus noise
intensity curves for several  $\beta _{L} <1$  and the corresponding
amplitude Fourier spectra of the output signal normalized by the Fourier
spectra of the input signal thus showing the spectral amplification
$k(f)$. It is seen that, for a sine signal of small amplitude ($a=10^{-3}$), the system response remains linear even for small
$\beta _{L} =0.1$, which is indicated by no sign a of third harmonic in
the output spectrum (even harmonics are absent due to the potential
symmetry). The latter case corresponds to the millikelvin temperature
range for real devices. Although it is obvious that the detected
spectrum is clearer at lower temperature because of a smaller noise
background, additionally the signal gain also turns out to
be high enough at $\beta_{L}=0.1$.

\begin{figure}[h]
  \centering
  \includegraphics[width = 0.7\columnwidth]{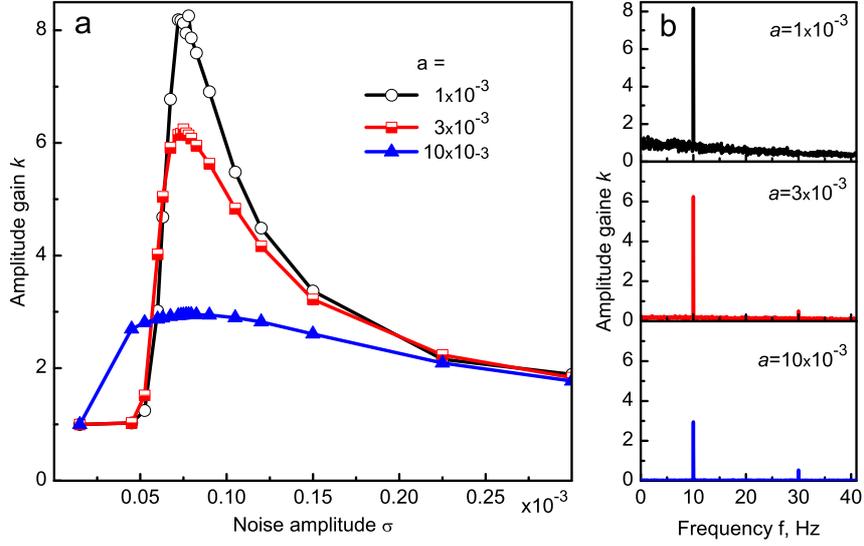}
  \caption{(Color online) (a) The amplitude gain of sine signals with
  various amplitudes $a$ in an ScS RF SQUID versus the noise amplitude
  $\sigma=D^{1/2}$. (b) The spectral amplitude gain $k(f)$ for the same $a$
  as in panel (a) and noise levels corresponding to the gain curve
  maxima in panel (a). The signal frequency $f_{s}=10$ Hz, the parameter
  $\beta_L=0.1$}. \label{fig5}
\end{figure}

     The effect of degradation of stochastic amplification in an ScS
SQUID with signal amplitude increase is shown in
Fig.\,\ref{fig5}. The higher the signal amplitude, the smaller the
signal gain, while the third harmonic (and other odd ones) in the output
Fourier spectrum becomes visible for $a=3\cdot 10^{-3}$ and $10^{-2}$
(even harmonics are absent because of the potential symmetry), thus
the amplification becomes markedly non-linear. Since signal-to-noise
ratio (SNR) enhancement in the output signal is hardly expected for
moderate-to-subthreshold signals on the background of rather weak
noise (associated with small $\beta_{L}$) \cite{Hanggi}, linear
amplification is more suitable in this case. Therefore, the weakest
signals are stochastically amplified by an ScS SQUID most effectively.

\begin{figure}[h]
  \centering
  \includegraphics[width = 0.7\columnwidth]{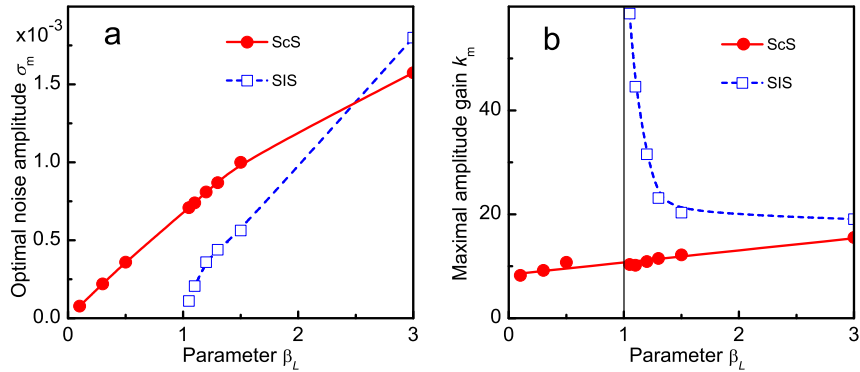}
  \caption{(Color online) (a) Maximum gain $k_{max}$ and (b) optimal
  noise amplitude $\sigma_{m}$ for ScS and SIS RF SQUIDs vs.
  parameter $\beta_{L}$. The signal frequency $f_{s}=10$ Hz and
  amplitude $a=0.001$.} \label{fig6}
\end{figure}

     The maximum stochastic gain for a weak ($a=0.001$ )
low-frequency ($f_{s} =10$ Hz) sine signal in both types of SQUIDs is presented in Fig.\,\ref{fig6}a versus the main SQUID
parameter $\beta _{L} =0.1-3$. The formal divergence of the signal gain
obtained for the SIS SQUID at $\beta _{L} =1$  will be smeared by noise
in real experiments. Besides, as an additional analysis shows, the
non-linear signal distortions drastically rise and the dynamic range
narrows in the region in the vicinity of $\beta _{L} =1$. For the
ScS SQUID, the dependence of the signal SR gain on the main
parameter  $\beta _{L} $  has no distinctive features within a wide
range of  $\beta _{L} $  including  $\beta _{L} <1$. The narrowing
of the dynamic range and rise of the non-linear distortion  is
observed at $\beta _{L} <<1$  similarly to SIS SQUID-based amplifiers near
$\beta _{L} =1$  due to a vanishingly small potential barrier.
Fig.\,\ref{fig6}b presents the optimal noise levels where maximum
gain is reached as a function of the parameter  $\beta _{L} $. As
expected, the optimal noise levels depend mostly on the height of
the barrier between the two metastable current states. It follows from
the obtained results that in the small signal approximation when the
response is supposed to be linear, SIS SQUIDs should be used as SR
amplifiers at  $\beta _{L} \ge 1$, while ScS SQUIDs are suitable for
small critical currents and/or inductances associated with flux
qubits, that is for  $\beta _{L} <1$.

\section{Conclusion}
\label{concl} In this work the noise-induced stochastic
amplification of weak informational signals at low temperatures
$T<<T_{c} $  in RF SQUIDs containing ScS contacts (QPCs) is
considered. It is shown that SR amplification of weak sine signals
emerges at any, vanishingly small, value of the parameter $\beta _{L} $.
This is due to an unusual shape of the potential barrier between the two
metastable states with a singularity at its top and always finite
height. It should be noted that there is no noise-induced
re-normalization of the potential energy of an ScS SQUID because the
noise is band-limited. This justifies the use of the
zero-temperature approximation.

Taking into account quantum corrections to the decay rate of the metastable
current states in SR \cite{Grifoni} can lead to essential
modification of the dynamics and rise of SR gain. For example, as reported
in the paper \cite{Omelyan}, under some conditions the
presence of noise could enhance the quantum correlation in
superconducting flux qubits. With temperature rise up to  $T_{c} $,
the SR dynamics of an RF SQUID with a QPC will change due to the temperature
dependence of the potential,  $U^{ScS} (T)$  \cite{Khlus}, tending,
apparently, to that of an SIS SQUID.

     It is worth noting that a discontinuous ("saw-like") current-phase
relationship at  $T=0$  is also a characteristic of other types of
Josephson contacts with direct conductance, e.g., the 4-terminal
microbridge junction and the superconductor-normal metal-superconductor
(SNS) junction, which results in a singularity on top of the
barrier of the potential for such junctions \cite{Ouboter,Kulik}, and
hence their stochastic dynamics should be similar to the behavior of an
RF SQUID with the considered ScS contact.

     In addition, we would like highlight one important feature of
SR. Even in the case when the SR effect in SQUID is considered as
"stochastic filtration" \cite{Klimont}, and no enhancement in the
signal-to-noise ratio is anticipated as compared to its "input" value
\cite{Hanggi}, the SR effect has an almost self-evident advantage over
other amplification methods because it works directly inside the
sensor, thus providing a kind of "first aid" to signal detection
that we could call "Just-In-Place Amplification" unlike widely
spread "On-Chip" technical solutions where amplification is carried out in a separate unit situated near the sensor on a common
substrate.\\

\textbf{Acknowledgements}\\

     The authors acknowledge Dr. A.A. Soroka for helpful discussions.\\

%\newpage

\end{document}